# Fragments of the Past: Curating Peer Support with Perpetrators of Domestic Violence

Rosanna Bellini

Open Lab, School of Computing, Newcastle University, r.f.bellini@newcastle.ac.uk

Alexander Wilson

Open Lab, School of Architecture Planning and Landscape, Newcastle University, alexander.wilson@newcastle.ac.uk

Jan David Smeddinck

Open Lab, School of Computing, Newcastle University, jan.smeddinck@newcastle.ac.uk

There is growing evidence that digital peer-support networks can have a positive influence on behaviour change and wellbeing outcomes for people who harm themselves and others. However, making and sustaining such networks are subject to ethical and pragmatic challenges, particularly for perpetrators of domestic violence whom pose unique risks when brought together. In this work we report on a ten-month study where we worked with six support workers and eighteen perpetrators in the design and deployment of Fragments of the Past; a socio-material system that connects audio messages with tangible artefacts. We share how crafting digitally-augmented artefacts - 'fragments' - of experiences of desisting from violence can translate messages for motivation and rapport between peers, without subjecting the process to risks inherent with direct inter-personal communication. These insights provide the basis for practical considerations for future network design with challenging populations.

## 1 INTRODUCTION

*"Desistance should not be seen so much as an ongoing event or state, but rather as … an ongoing work in progress … the going is the thing"*, Fergus McNeill and Shadd Maruna, Giving Up and Giving Back (2007) [49].

Perpetrators of domestic violence frequently require long-term support from peers and professionals to desist from using abusive and harmful behaviours against others. Domestic Violence Prevention Programmes (DVPPs) are one way of providing perpetrators a pathway to non-violence by re-educating attendees on the unacceptability of using violence against their victim-survivor(s). In doing so, group-work programmes report improved health and social care outcomes for those involved by cultivating non-coercive relationships between attendees for mutual support and responsibility for change [50]. However, the loss of positive peer groups, for instance at the end of a DVPP, has been identified as a significant risk factor for the reuse of abusive and harmful behaviours towards victim-survivors [1, 54, 55]. Peer support networks following the conclusion of behaviour change interventions are one attempt to extend the life of these relationships as these require peers to share emotional, social or practical advice, whether located in-person or hosted online [38, 60]. Nonetheless, peer support networks, particularly those facilitated through online platforms, are subject to their own challenges. These challenges include encouraging adequate participation from group members or applying special attention to the ethical dimensions of monitoring discussion around harmful topics. As such, work is required to understand the crucial stage of completion of a DVPP to identify further opportunities for violence prevention and better ensure the long-term safety of victim-survivors.

In this work, we report on a ten-month study we conducted in collaboration with a domestic violence service provider to co-design a safety-focused, moderated peer-support network between two groups of male perpetrators in DVPPs. We present our analysis of five design workshops, a custom deployment of the JigsAudio system [83]: *Fragments of the Past* for asynchronously translating supportive messages, and how this deployment was received by our participants. Through this work, we discuss how digital artefacts that remain fixed across time that we term in this work as possessing *temporal permanence* helped us navigate the challenges of engagement and risks inherent to exchanging safe guidance on violence prevention. Our study design was shaped by these practical and ethical considerations in implementing our research efforts, for which the aim was to investigate the following research questions:

*RQ1. How can digital peer-support be configured to safely accomodate perpetrators of domestic violence after the conclusion of a domestic violence perpetrator programme?*

*RQ2. How might digital peer-support processes be designed to address ethical concerns around potential negative feedback loops for perpetrators of domestic violence?*

We thereby contribute to the growing discourse in HCI on designing with professional services that work with people who use violence in their intimate relationships in the following ways: (1) we identify the facilitators and inhibitors for peer support exchange between users of harmful behaviour; (2) we provide a novel deployment of an asynchronous peer support network *Fragments of the Past* with perpetrators of domestic violence; and (3) we outline how the temporal permanence of artefacts can act as a design approach to addressing the organizational and ethical challenges around synchronous communication.

## 2 BACKGROUND AND RELATED WORK

Peer support is a process where people who share common experiences or face similar challenges come together to give and receive help based on the knowledge derived from shared experience [64]. Importantly, peer support can positively benefit both the person receiving support and can make the provider feel valued, needed and included [68, 69]. Peers can also influence the behaviour of other group members that can be pro-social or harmful [19, 25], or be a combination of the two [75]. Moreover, peer support can transcend or extend traditional social care delivery settings, making it a viable option to reach minority or marginalised populations [71] including people impacted by domestic violence [17, 36]. Such a process can be delivered through one-on-one support by a trained peer, team-based support, or peer-ran groups [23]; be organically occurring (i.e. a spontaneous connection between people) or structured by a professional provider [23, 50, 64]; and be mediated with or without digital technologies [59].

### 2.1 Peer Support Technologies for Reducing Harm

Technology-mediated communication has been described as a "*promising new direction*" in peer support [86], particularly for facilitating sensitive discussion with under-served or at-risk social groups. HCI researchers and technology designers have successfully expanded our knowledge of how people might use unmoderated communities and social network sites to support behaviour change away from using harmful behaviour(s) [47, 67]. Online peer support has grown in popularity due to the ease of access, flexible participation, the ability to maintain a degree of privacy, and a decreased communication apprehension due to reduced social context cues [9, 11]. However, such investigations have also detailed the challenges within (often unmoderated) online peer support. Many online forums that facilitate peer support can have uneven 'bursts of activity' and lack consistency of engagement over time [44]. When stigmatizing behaviours are being disclosed, moderators or a collection of users in online support groups may further reinforce harmful behavioural, cultural and social norms for an individual user [16, 62, 77]. If this influence is extremely successful, a person's real-life relationships can decline in quality, at worst subverting the positive social potential of support groups [52]. These risks are heightened with perpetrators of domestic violence due to many individuals being well-versed in the social manipulation of others [35] and the institutional systems that seek to hold their abuse to account [34, 55].

To mitigate the challenges for online support groups, some scholars have explored a hybrid-approach to better understand how technology can leverage existing face-to-face peer support networks and social structures [20, 38, 59]. In the community-driven approach of *Circles of Support and Accountability* (CoSA), a circle of trained volunteers connected via a digital network work with sex offenders to minimise alienation and support reintegration after incarceration with the aim of preventing sexual reoffending in their community [84]. Nicholson et al.'s approach in [58] provides a conceptual framework for training community peers to act as 'guardians' to disseminate preventative strategies to their community to increase cyber resilience. Heyer et al. have also recently highlighted strategies for social computing tools to support dyadic mentorship for recovery from alcohol or drug dependency [38]. However, prior research has yet to investigate how technology might support and expand safe interventions for supporting the desistance of abusive behaviours towards others, despite there being calls to do so [30, 66].

### 2.2 Responsibility for Change beyond a DVPP

*Domestic Violence Perpetrator Programmes* (DVPPs) are a preventative strategy against patterns of domestic violence. While the content of a DVPP may vary, most programmes within the UK aim to stop abuse through challenging the pro-violence attitudes of the perpetrators [26], and they employ cognitive behavioural techniques to educate them on non-violent alternatives [5, 15]. Most DVPPs require attendance for two hours a week (M = 112 min, SD = 64) in a group therapy format for between 24 to 48 weeks (M = 26, SD = 17) [34]. This combination of challenging harmful attitudes, while providing individuals with tools for behaviour change is emotionally demanding for many participants enrolled in interventions [27, 28, 55]. As such, most programme evaluations promote an understanding of desistance from violence as neither a linear progression, nor the immediate termination of abuse, but a complex, dynamic pathway that gradually unfolds over time [13, 49, 78]. This causal process can require continuous engagement over long time-periods, or even a lifespan, to change to a non-violent identity.

Morran reports there is a stark scarcity of post-programme interventions or resources for perpetrators to continue "*maintaining the momentum of change*" when dealing with wider challenges within their lives [55]. Attendance in a DVPP can act as a protective factor against abusive behaviour [82], through pro-social bonds between attendees and facilitators through mutual support and accountability [50]. Inversely, the loss of social support is a strong risk factor for relapsing into the perpetration of domestic violence [1, 85]. While there are concerns for collusion [72] and risk-escalation [63] in re-grouping groups of perpetrators together without moderation, studies indicate the opposite for moderated spaces [2]. However careful moderation has high time and cost investments, and there are still considerable challenges around sustained engagement with in-person and online communities. As such it is crucial to explore how to build and sustain interventions for perpetrators to provide and receive peer support on the dynamic pathway of desistance. This is in line with Bellini et al.'s appeal to "*provide physical, virtual and social spaces of negotiation for perpetrators*" [6] to realise the extent of their abuse, their duties as a non-violent person and act independently of violence prevention services.

### 2.3 Asynchronous Communication and Tangibility

When emotive conversations take place, many people enrolled in harm-reduction interventions express a preference for face-to-face, real time communication between peers either in-person or via video conferencing [3]. These modes of communication require precise synchronization in time and location of many individuals, and thus can be difficult for people with inflexible work-schedules, lack of access to transport or who might encounter other technological barriers to attend virtual sessions – all challenges identified for perpetrators of domestic violence enrolled on DVPPs [40]. As a result, scholars have praised the role of asynchronous communication in the facilitation of peer groups as these methods can provide flexibility to delivery and moderation [61]. By attempting to reduce the digital proficiency barrier that some participants face, many scholars have long called for designers to move away from screen-based, text-based communication such as a dependency on social media channels, in order to explore alternative yet meaningful design processes and practices [31]. This is because such work identifies that the value in the *relational* translation of peer support and the expressivity of giving and receiving of such support or care can look distinctively different depending on the context [74]. Such work has examined how collectives such as friends, couples or family members may share digitally-mediated social objects (physical objects for which symbolic value lies in how they represent social relationships) as a means of communication [41]. Gift-giving and memorializing everyday memories are two ways that HCI has explored the exchange and reciprocity involved between peer relations [56, 80]. While asynchronous communication through artefacts is an established research space, this work is the first to our knowledge to focus on how such entities can be leveraged as motivation and communication for peer support in behaviour change.

### 2.4 Making for Representing Self and Change

Goodman et al.'s work argues that the process of making inherently stimulates memories of people, relationships, activities and emotions [33], by enabling the maker to communicate something about themselves through the things they create [87]. Importantly, making can serve to facilitate education on challenging topics by providing an abstract medium in which to focus attention while engaging in social conversation. Changing Relations' *Men's Voices* project [88] is such an example, where crafting was used to gather testimonies of young men and boys' experiences of contemporary masculinity, and as a means to teaching healthy relationships for domestic violence prevention. The process of making can also have a positive impact on participants by providing a space for reflexivity and positive enforcement through a sense of control over the artefact being created, while prompting a sense of competence and a social connection between people. [22, 70, 73]. This can be evidenced in Thieme et al.'s work within a women's secure psychiatric unit to form *Spheres of Wellbeing* to support the learning and vital practice of Dialectical Behavioural Therapy (DBT) that included the promotion of mindfulness and the tolerance of emotional distress [76]. Clarke et al. used photo-sharing and participatory storytelling between a group of victim-survivors at a women's centre to facilitate identity-sharing and identity-construction after leaving an abusive relationship [18]. Memorialisation has also been explored through artefacts and processes, notably Moncur et al.'s *Storyshell* to commemorate personal bereavement [53], and Durrant et al.'s study into the human values present within archiving historic accounts of political violence [29].

In summary, there exist challenges for facilitating synchronous, yet moderated peer support networks between perpetrators of domestic violence for supporting desistance from abuse. Still, the withdrawal of peer support at the conclusion of DVPPs has been identified as being a significant risk factor in the recidivism of abuse. To constrict this process could mean that there is a lack of accountability and responsibility for perpetrators as they face challenges in their own lives. As such, it is important to explore how asynchronous modes of communication might be able to facilitate such support, and what role socio-technical services can play in avoiding the challenges faced by social-media based systems.

## 3 STUDY DESCRIPTION

For this study the research team worked in close collaboration with a national vulnerable people and domestic violence charity <Stop Violence> at their regional hub <Cloudside> in the North of England, UK[1]. <Cloudside> is made up of a large staff team who deliver a portfolio of interventions for both victim-survivors and perpetrators of domestic violence, working with local voluntary and statutory organisations to coordinate the risk management of perpetrators in their region. After several group meetings with the lead author, six members of the <Stop Violence> team decided to explore how post-programme service users could act as peers for early-programme service users. The staff saw this process as resulting in two positive outcomes: early-programme service users would receive guidance (moderated by <Stop Violence> staff) from post-programme service users to encourage engagement in the behaviour change process, while post-programme service users could adopt new roles for responsibility and sustain healthier forms of communication [6, 50]. Two groups of programme participants were subsequently identified as being suitable for this investigation (**Error! Reference source not found.**) located across <Stop Violence>'s delivery groups: the Norstone Group (10 male perpetrators) and the Grandvens Group (8 male perpetrators). Each group contained a mixture of self-referrals and men requested to attend via court order through the involvement of law enforcement. So as to not bias the researcher's approach to participants (e.g. preferencing to work with voluntary perpetrators rather than mandatory [27]), referral information and a perpetrator's prior offenses were withheld by the support organisation. Each man was notified that participation in the study would not impact on their course progression with the exception of disclosing safeguarding concerns that could violate the programme integrity.

[1] All data has been appropriately de-identified and anonymised to protect the identities of our research partners and participants. We share further details of our approach in 3.3 Data Analysis.

### 3.1 Ethics

Both perpetrators and victim-survivors are classified as at-risk and vulnerable populations. As there are a number of ethical considerations that are especially acute within studies on violence and abuse [28, 81], we share our steps to inform safe and ethical practice here. Within our research, we prioritised to work with <Stop Violence> as the organisation has received a quality assurance accreditation by the national sector leader for safety-focused practices for working with victim-survivors and perpetrators. Each facilitator (F1 – F6) included in our research has extensive experience in working with individuals who harm. Facilitators F1, F4 and F6 assisted in co-designing the research sessions to identify risks, challenges or threats to safe-guarding for victim-survivors, staff, perpetrators and the research team. After looking over our research plan, each facilitator judged our activities unlikely to cause an undue escalation of risk to service users (including victim-survivors) and reminded the research team of existing safety procedures in the event that the research team or a participant needed additional support, advice or care.

Our participants were a representative sample of the risk profile (standard-medium) of perpetrators enrolled on DVPPs [24]. Written consent forms were used at the start of each major study stage and verbal consent was also sought by each participant before research could commence. Of the participants who agreed to participate in our study, roughly one third of the perpetrators were in new relationships, one third outside of a relationship, with the final third still partnered with their victim-survivors. Each current or ex-partner of perpetrators enrolled on the DVPP – and by extension our study – is offered support through an integrated safety service to coordinate care, communication and ensure safety checks are conducted on victim-survivor wellbeing. Irrespective of the relationship status with the perpetrator, each man within our study was asked to refer to victim-survivors by their first name in conversation in an attempt to respect and humanise them in discussion. While our work sought to explore the positive formation of peer relationships, facilitators also sought to challenge negative collusion between our participants should it occur by intervening and offering one-on-one work after the research study had concluded.

Finally, the lead author has years of experience in working to design, deploy and evaluate technologies with support organisations who work with perpetrators of domestic abuse [5, 6, 8]. While it is reportedly rare for perpetrators to pose a physical risk to researchers (as most do not use violence outside of their relationships [37]), she also possesses appropriate training for dealing with extreme and difficult behaviours. To mitigate the impact of being targeted by the perpetrators outside the session, she disclosed only her first name and did not share any identifiable details or contact information. Finally, as working with any topic related to violence and abuse can be emotionally and mentally challenging [14], she also has a self-care plan in place for dealing with the effects of vicarious trauma.

### *3.1.1 Ethical Approval*

The research team submitted an ethics application to the lead author's research institution that outlined the project aspirations; to identify an existing or a bespoke piece of technology that would permit the Norstone Group (Group N) to provide peer support, moderated by professional facilitators (F4, F5, F6), to members of the Grandvens Group (Group G). Ethical approval was granted to carry out the research under the condition: "*that participants should not have any opportunity to contact with one another*", referring to the communication between Group N and Group G. The <Stop Violence> team outlined that this ethical decision disallowed the standardised delivery of a DVPP where attendees to engage with each other as they are encouraged to do so within a moderated, group-therapy format. Additionally, the team outlined that this verdict differed from widely accepted ethical practice for research within contexts of domestic violence: that participants were active agents in the process; that they could make choices about their lives and be afforded opportunities for the positive experiences and impacts of the research [42]. Nevertheless, this condition was non-negotiable as the ethics committee clarified that the institution could be in no way legally responsible for the risks incurred by the project (a response discussed in prior HCI work [12]). As such, the research team and the staff identified that a 'live' synchronous network was not an option for this project. Built on a motivation to work within this context, and in acknowledgement of the complexity for sustainability of traditional supportive group settings [39, 44], the research team and <Stop Violence> came to understand the verdict as a design challenge; *how could digital peer support be facilitated without permitting direct communication for perpetrators of domestic violence?*

| | |
|---|---|
| **10 Perpetrators of Domestic Violence: Norstone Group (N1 – N10)** | |
| Age (years) | 20 – 65 Average: 41 |
| Sex | Male: 8 Female: 0 |
| Risk Level | Standard[2]: 7 Medium: 3 |
| **8 Perpetrators of Domestic Violence: Grandvens Group (G1 – G8)** | |
| Age (years) | 24 – 59 Average: 38 |
| Sex | Male: 8 Female: 0 |
| Risk Level | Standard[1]: 6 Medium: 2 |
| **6 <Stop Violence> Staff (F1 – F6)** | |
| Age (years) | 27 - 61 Average: 43 |
| Sex | Male: 1 Female: 5 |
| Professional Roles | Head of <Cloudside>: 1 (F1)<br>Case Manager / Worker: 4 (F2, F3, F4, F5)<br>Group Facilitator: 1 (F6) |

Table 1: Participant Demographics of Age, Sex, Risk Level (Perpetrator) or Professional Role (<Stop Violence>)

[2] Perpetrators are assessed using the DASH Risk Checklist that evaluates potential risk factors of a perpetrator to a situation. Standard refers to a 'low' risk of immediate threat to harm or murder while Medium can indicate other risk concerns such as history of violence, pregnancy of victim-survivor and so on.

### 3.2 Study Design and Participants

Our study ran over ten months with four stages: 1) five design workshops with attendees of Group N to design an asynchronous support network activity; 2) the design of a digital system *Fragments of the Past* (FoTP)*;* 3) the deployment of FoTP with Group N; 4) a structured reflection and commentary with Group G on their use of FoTP. Our first study stage used five design workshops to gain an understanding of how men in Group N were providing peer support to each other, and to what extent technology could play a role in facilitating a novel network. The second stage describes how the workshop findings were used to transform an existing piece of technology (JigsAudio [83]) into a socio-material peer support network activity for Group N in the creation of digital artefacts for representing important moments change to non-violence, their so-called "fragments of the past" (FoTP). The third stage describes the deployment of FoTP with Group N and a representation of these created fragments. Finally, we conclude with the fourth stage on the results from a three-hour structured critical reflection with Group G on Group N's fragments to capture their thoughts on the process of receiving support through this mechanism. As the results of each stage directly informed the design of the next stage, our research paper is structured as such.

### 3.3 Data Analysis

Each research stage was audio recorded for a total of 19 hours and 26 minutes that was partially transcribed for relevance with confidential conversations removed. All qualitative data were analysed using a constructivist-realist Grounded Theory approach [21, 32]. This required the first author to code line-by-line in a round of open coding that produced a qualitative codebook of 81 codes that was shared and agreed upon with the research team. The second stage of axial coding produced 13 concepts before a final round of selective coding identified 6 categories that we present across the stages of this paper. The quantity of our codes, concepts and categories are within the Lichtman's ratio of 16:3:1 for qualitative analysis [45]. For the visual artefacts in Stages 2 and 3 we used a Visual Grounded Theory Methodology (VGTM) by Mey et al. by following their six procedural steps as a framework of orientation toward the investigation of visual form [51].

### 3.4 Stage One: Design Workshops with the Norstone Group

In this stage, the lead author worked with ten perpetrators of domestic violence that were arriving at the end of their DVPP together. The research team were aware that preparing participants for design work is an acute challenge in sensitive settings [10, 46]. Therefore, a series of design workshops were conducted to gradually build sensitisation for Group N to understand themselves as peers to others. The workshops took place between October and December 2018. These sessions accommodated the materials from completed DVPP modules to discuss relevant supportive advice, included a review of the technologies suggested by the men, and provided a space to reflect on the group process together. In line with the standard format of DVPPs, design activities were set as homework for the group to complete at home between sessions to permit further reflection on care and support. Below we provide an example of a homework activity pack *Making Connections* (**Error! Reference source not found.**) and an interaction design activity in a workshop session *Three's a Crowd* (Figure 2).

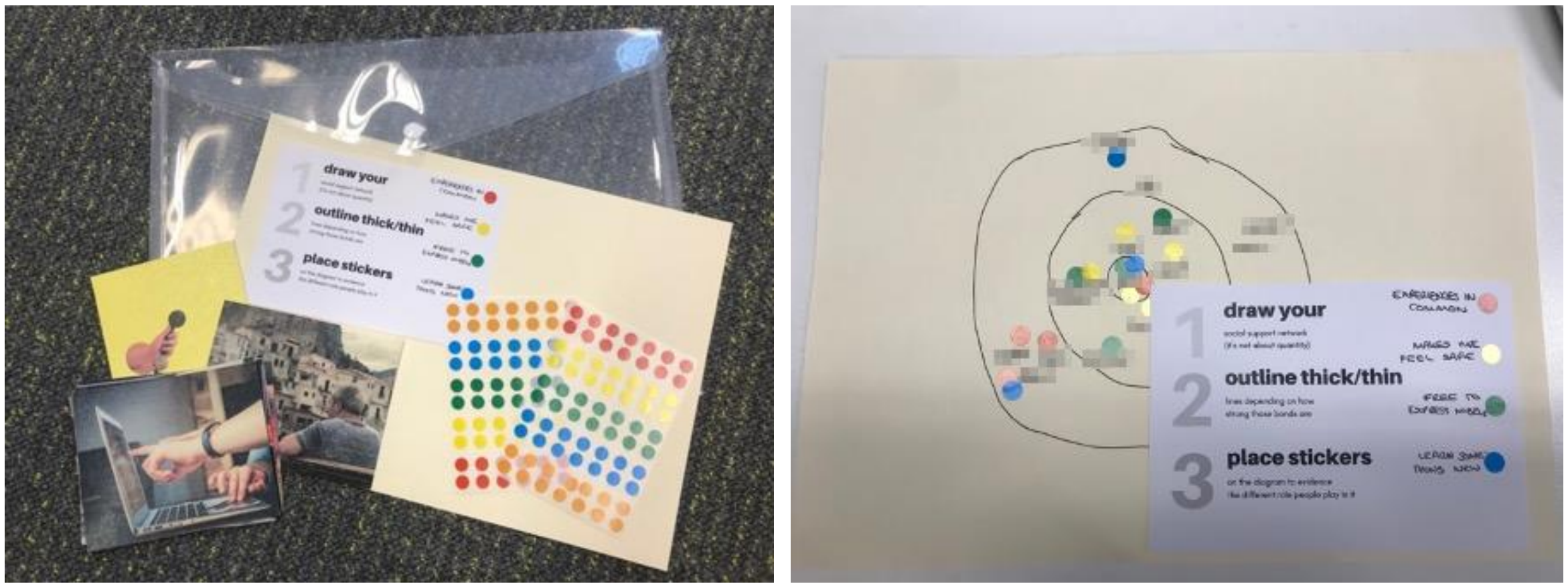

Figure 2: Making Connections pack Incomplete [left] and Complete [right]

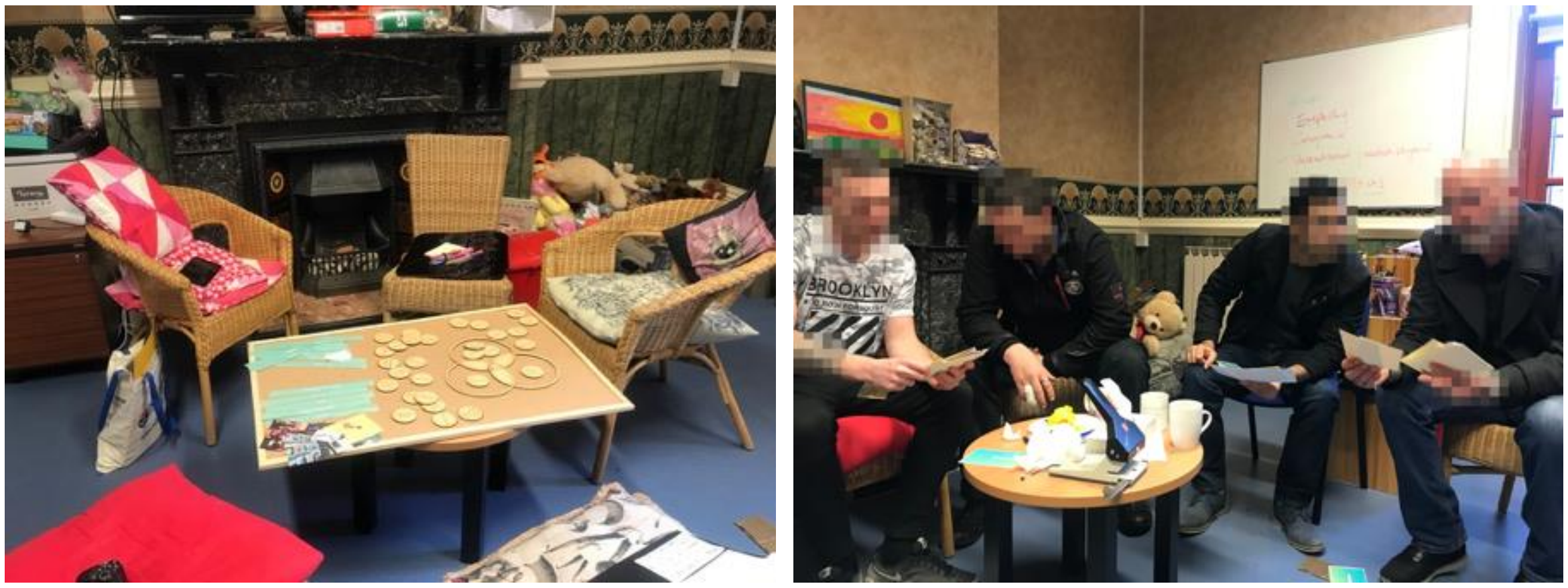

Figure 1: Three's a Crowd Set Up [left] and Four Participants Completing the Activity [right]

| Category | Quantity | Values |
|---|---|---|
| What | 11 | Smartphone Application, Podcast/Radio, Website, Social Media, Playlist, Instant Messaging, Digital Art Installation, Blog, Photo Album, Email, Text Messaging Service |
| Why | 9 | Self-Reflection, Communication, Motivation, Education, Entertainment, Relaxation, Storytelling, Self-Care, Support |
| Qualities | 14 | Helpful, Informative, Creative, Inspirational, Mindful, Humorous, Tolerant/Non-Judgmental, Easy-to-Read/Listen To, Simple, Fun, Interesting, Realistic, Emotive, Genuine, Truthful |

Table 2: Three's a Crowd Token Descriptors

For *Making Connections* (**Error! Reference source not found.**), each man was tasked with completing a pack that asked them to represent their personal support network, how strongly they connected with people within it and the different role those people played within the network. Using the material from the completed homework packs, a second design activity consolidated our understanding on the feasibility for a digital asynchronous peer support network between Group N and Group G. For *Three's a Crowd* (Figure 2), participants were asked to identify 'what' technology the network would use, 'why' it would be useful and what five 'qualities' would be most important to the system. Each category was represented as a series of wooden tokens that could be placed within a Venn diagram to encourage reflection on the relations between the people, qualities and objects for a final design (Table 2). The tokens reflected what the participants described from their own peer support network (*Making Connections*), and the technologies used in maintaining them. This activity had three rounds, entailing that a minimum of three peer support technologies would be created.

### 3.5 Stage One: Findings

The five design workshops provided the research team with an in-depth insight into the ways that our participants understood their desistance from abusive behaviours and their desire to support other perpetrators starting a DVPP. We now share two categories on providing peer support through digital technologies that we were able to identify from our data; *balancing parts with wholes* and *mixing the digital and non-digital*.

#### *3.5.1 Balancing Parts with Wholes*

Many participants disclosed that at the start of the DVPP, a large challenge for becoming engaged in changing their behaviour was based around anxieties for how they would be judged by facilitators and other perpetrators. N2 shared that it had been intimidating for him, in his words, to be seen as *only* the sum of his abusive behaviour, particularly through digital case files:

> "… there's no way around excusing what I did, none whatsoever, I was a nasty piece of work like. Until I got to <Cloudside> I felt like no one saw me as me and not my actions on my record … the facilitators here, they tried to work out what was going wrong somewhere." (N2, Perpetrator)

As a way of managing this discomfort, many participants were interested in depicting themselves via an avatar or a collection of possessions to represent a digital past self who used violent behaviours, and a present self who had stopped using abuse. This caused tension between the facilitators and some participants who stated it could be unhelpful to compartmentalize abuse behaviour with a "*bad version of themselves*" (F4, Facilitator) instead of owning up to their use of violence. One man agreed this separation could minimize their responsibility for violence by claiming it "*wasn't really them*" (N8, Male Perpetrator). The group discussed at length that the comparison to prior bad behaviour was an important motivator for them to continue to desist from abusive behaviours as they saw it as evidence that change was possible through a DVPP. As such, the group identified there was a delicate balance between disowning and owning past behaviours. One way of rectifying this challenge was by participants identifying what *parts* of themselves, such as memories (represented via photos) or thought processes (represented via blog posts) they wished to work on, were proud of or wanted to keep the same. This was demonstrated through the men repeatedly gravitating towards the *Blog*, *Photo Album* and *Playlist* tokens in *Three's a Crowd* (Table 2).

> "… that's what the programme is meant to do, understand yous [yourself]… at the start you're not honest with yourself, but you gradually open yourself up to see what pieces you're made up of" (N6, Perpetrator)

#### *3.5.2 Mixing the Digital and Non-Digital*

Many participants were curious and enthusiastic to describe their own speculative technologies that could play a role in providing support to new-starters and the scenarios in which they could be used. As some participants did not own a smartphone (NG3, NG9), the group were resistant to advocate for screen-based technologies out of concern of excluding these participants. This entailed that many participants first sketched out an idea about what kinds of support they wished to share (aided by the homework activity *Making Connections*) before adding additional layers of digital elements and double-checking that the group were familiar with using these technologies. This permitted the participants to think about an idea for support first without being confronted or frustrated with the technology to begin with, or inadvertently, for the researchers to impose their desire for the process to be digital in nature [4]:

> "… if you introduce it as 'technology' the men can feel like they're on the back foot … but if it's technical but not 'scary', say combine it with things they are familiar with you'll get past that initial resistance" (F1, Head of <Cloudside>)

Many participants also expressed enjoyment from the tangible and creative nature of the activities (such as *Three's a Crowd*) within the design workshops. By this, participants shared an appreciation for the opportunity to express themselves through craft that in turn extended the learning from the behaviour change sessions:

> "So, you learn problem solving skills in the group, thinking of another way to not be mean, nasty … you gotta get creative and I'd like that quality to be [at] the centre of whatever we make." (N5, Perpetrator)

Some participants identified that using familiar materials to be creative with, such as photographs and audio recordings that are ubiquitous in everyday use, could ensure that a digital support system could be both accessible (not put someone "*on the back foot*" (F1)) and creative. <Stop Violence> staff were especially interested in how the role of creative practice could also be used as a channel to engage the men in challenging conversations about violence in the future.

## 4 STAGE TWO: FRAGMENTS OF THE PAST

At the conclusion of the design process, Group N had finalized their design of creating and sharing digitally-enabled artefacts to act as an asynchronous peer support network in accordance with the ethical dimensions of our work. This design involved a tangible, digital scrapbook, containing their so-called "*fragments of the past*" (N8, Male Perpetrator) including their stories of change, pieces of advice and supportive messages for Group G. These were shared through photographs and pre-recorded audio recordings activated by a digital playback button at the bottom of each page. While facilitators agreed the scrapbook was a positive way of providing peer support, they were concerned about <Cloudside>'s tightly restricted financial budget and the project timeline required to make the design a reality. Building on the findings of *balancing parts with wholes* and *mixing the digital and non-digital*, the lead author navigated this problem through introducing three mixed-media technologies that had previously been used in civic settings with vulnerable people could be repurposed by the group. These technologies were presented through a structured focus group with Group N and facilitators that included: *Gabber* a platform for distributed audio capture and participatory sensemaking [65], *JigsAudio* a tangible device that connects physical objects with an audio recording [83] and *Lifting the Lid* a digital probe that played back a pre-recorded message when it was interacted with [7]. Each technology was trialled via a run-through, before the lead author questioned how the group could see aforementioned *fragments of the past* work with the designs. After careful deliberation, the group decided JigsAudio would be a suitable technology for this task, as it could connect the digital media that they had been intending to put in the scrapbook with an audio reflection on their behaviour.

JigsAudio was developed as a technology that supports people sharing their experiences and aspirations of where they live in response to open questions. The device and method use a tangible hardware hub (**Error! Reference source not found.**) to connect drawing with talking through an embedded RFID connected to a physical artefact. Once a physical artefact has a recorded audio reflection attached to it, the artefact can be held over the physical system and the audio is replayed (e.g. over headphones). It is comprised of a Raspberry Pi, a microphone, a portable battery, and an RFID (radio-frequency identification) reader within a customised enclosure (Figure 3).

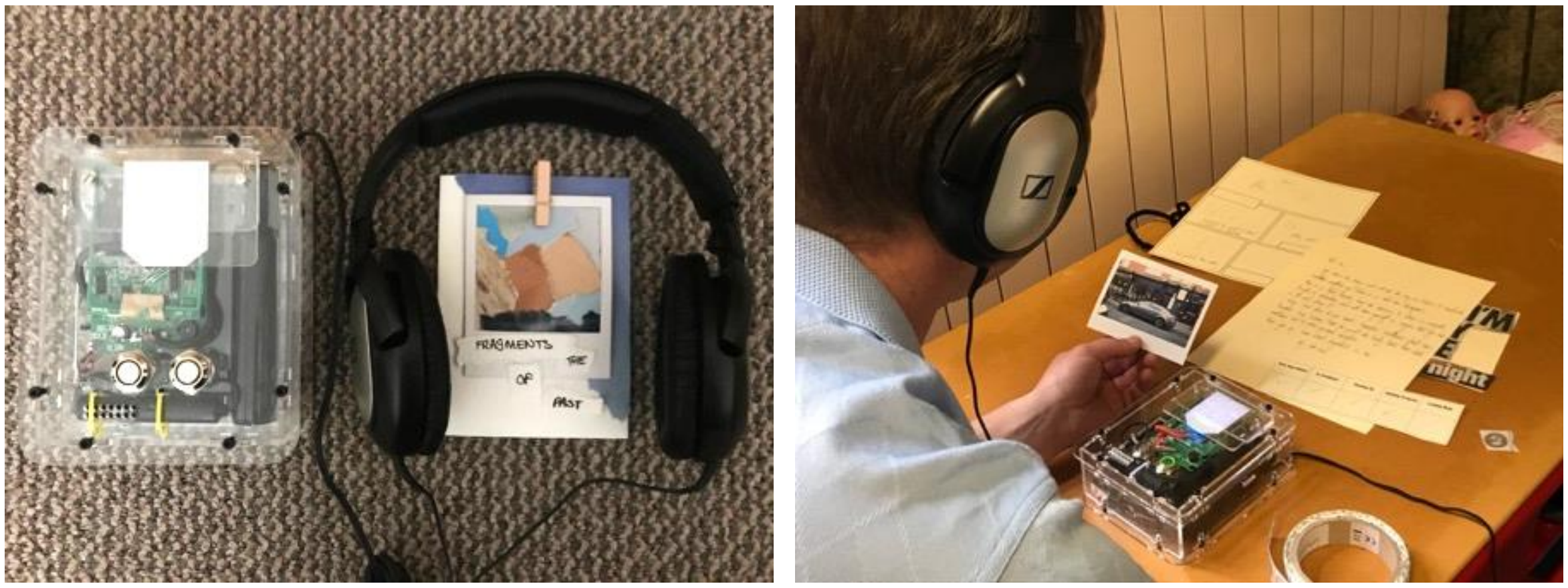

Figure 3: [left] JigsAudio System with FoTP Passport [Closed] and a Group G participant listening to a member of Group N's Fragments [right]

To use JigsAudio for *Fragments of the Past*, group members of Group N wanted to share their thoughts, feelings and experiences towards themselves and their behaviour across key moments on the DVPP. To do this, men requested to make a tangible representation of these key moments using craft materials (including card, newspaper, pipe cleaners, nuts and bolts) forming a 'fragment' of themselves from the past and attach at least one audio reflection with each artefact using an RFID sticker. Each man suggested using the craft material to complete a collection of crafted fragments: a comic; an abstract model; three polaroid photos (excluding names and faces); a letter to someone (including themselves) and a collage. This would result in a minimum of seven physical artefacts and audio recordings per person. After careful deliberation, five important moments along their journey towards non-violence were decided on by the group: *First Impressions*, *In Avoidance*, *Opening Up*, *Making Progress* and *Looking Back*. These moments were printed as destinations within a paper passport that each participant had to stamp off once they had crafted a fragment that corresponded with one of the five key moments (Figure 3). We identified that defining the boundaries of the activity appeared to reduce some participants' anxiety surrounding creativity as it provided direction to the creation of their physical fragments yet was flexible enough to remain thematically open. Importantly, the activity was also to ensure that the participant felt like they had "*done enough, that they have contributed something worthwhile*" through the completion of discrete steps within the activity [79].

## 5 STAGE THREE: CRAFTING THE FRAGMENTS

A four-hour group workshop was organized to produce the fragments to be shared within the FoTP network. To reduce disruption to the participant's schedules, the workshop was scheduled to coincide with a voluntary post-programme social meet-up for the Group N at <Cloudside>. Two facilitators (F5, F6) also participated with Group N on the activity to make the participants feel more at ease, though their fragments were not included in the analysis of this work on their request.

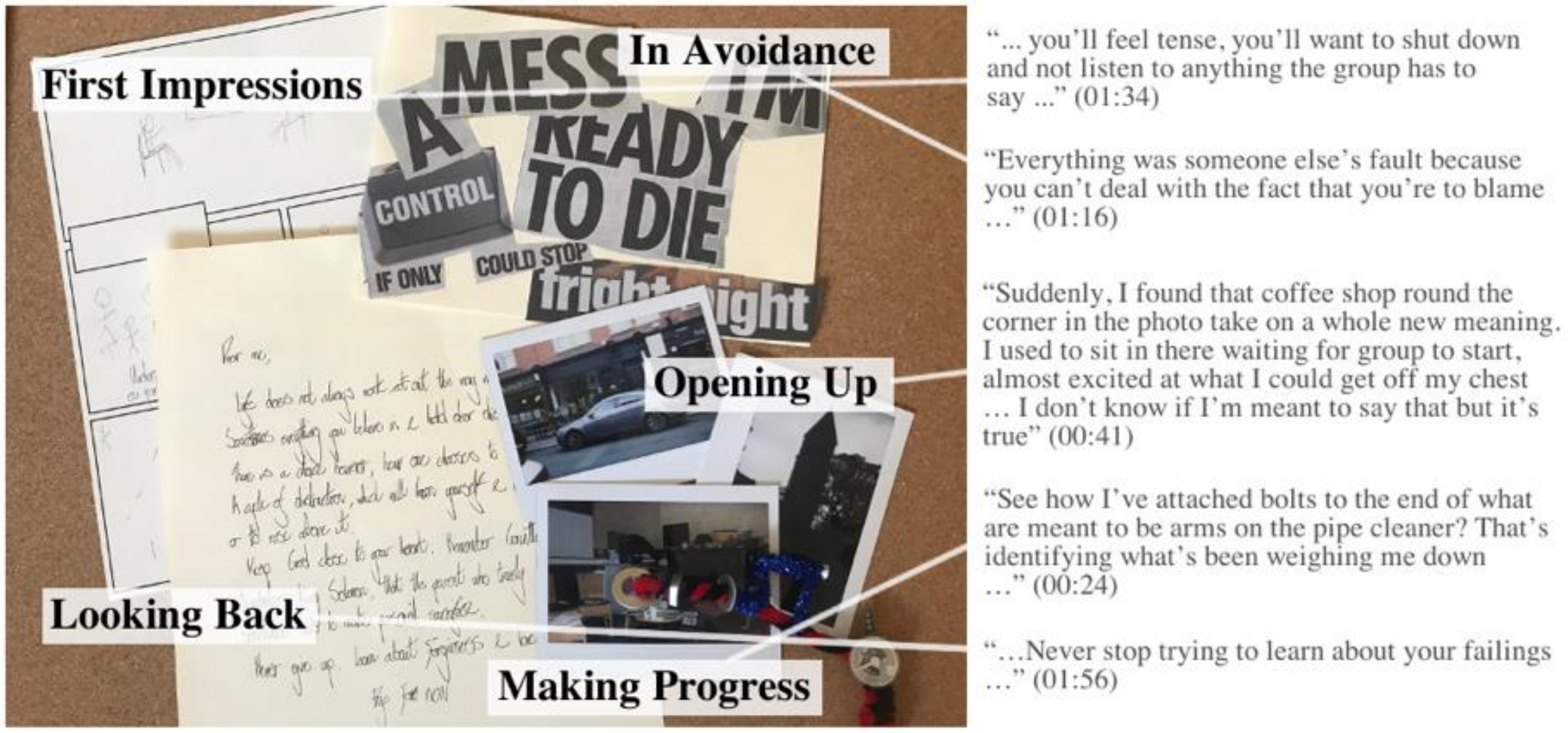

Figure 4: A Participants Finished Set of Fragments

### 5.1 Stage Three: Findings, Artefacts

Each group member completed at least seven fragments that represented key moments in their behaviour change journey. Across the four-hour session, 88 fragments were made by the Group N who recorded 85 audio clips associated with the fragments. One participant expressed anxiety about participating due to feeling intimidated by the variety of unfamiliar mediums to work with. In being responsive to his concerns the facilitators suggested he could participate in the process through channelling his skill for technical sketches and diagrams into fragments instead of the materials provided. We have included these fragments as 'Misc.' category in Table 3. Regarding the slight variation in the number of fragments per individual, some participants identified that there was more advice that they wished to share about their journey and thus crafted another. The small discrepancy between the audio recordings and physical fragments was occasionally down to some fragments taking an awkward physical shape (such as the abstract model) so that the RFID stickers were challenging to attach. We identified that the longer audio recordings for the comic, letter and the collage in comparison to the photos or model was due to each man explaining the complexity of each drawing, or some men opting to read their letter out loud. Participants particularly gravitated towards the polaroid photos due to their interest in representing important places or objects within their fragments – an interesting mirror of Clarke et al.'s findings that photo-sharing with victim-survivors also appreciated using *metaphor* for representing topics related to violence [18].

| **Fragments of the Past** | | | | | | | |
|---|---|---|---|---|---|---|---|
| | Polaroids | Comic | Letter | Collage | Model | Misc. | Total |
| Number of fragments | 36 | 11 | 13 | 10 | 12 | 6 | 88 |
| Number of audio clips | 33 | 11 | 14 | 10 | 10 | 7 | 85 |
| Average length of audio (mm:ss) | 00:34 | 01:49 | 01:53 | 01:16 | 00:38 | 00:55 | 01:03 |
| Total length of audio (mm:ss) | 18:42 | 19:59 | 26:22 | 12:40 | 06:20 | 06:26 | 90:29 |

Table 3: Number of Fragments and Audio Clips

**Stage Three: Findings, Themes**

Throughout fragment crafting process, the research team was interested in exploring; what kind of support information were shared by Group N, how was this visually and audibly represented using FoTP and how Group N felt about the process of crafting the fragments. Through our GT and VGTM analysis [21, 51], we identified two qualities in this deployment: *Audibly Augmenting Reality* and *Curating Identities*.

### *5.1.1 Audibly Augmenting Reality*

Each participant primarily chose to communicate emotive and encouraging guidance through their fragments, yet this was done in distinctively different ways through the audio recordings. We were able to identify three strategies for how participants understood the connection between their physical fragments and the supportive messages connected to them; that their fragments were *evidence* that behaviour change was possible; that they could *challenge* false narratives about DVPPs and could help to *explain* the complex thoughts and feelings at particular moments in the behaviour change process. Many participants identified that a significant challenge for them at the start of the DVPP was a lack of 'proof' that behaviour change was possible. As such some men positioned the fragments as *evidence* that it was possible to move on from and live a life desisting from the use of violent behaviours. One participant wrote a letter to himself addressing how he remembered he had thought about his use of violence:

> "believe me man when I say that it is possible to do something about you and your behaviour … I genuinely used to think this way this letter talks about myself and [name of victim-survivor] but I don't any more …" (N6, Perpetrator)

The most common approach was using fragments and the associated audio to *explain* and describe their feelings or thought patterns in key moments of a DVPP, sometimes by describing what particular colours or drawings meant to a participant (Figure 5):

> “the brown for me represents how shut off I was, I wouldn’t listen … I was inside my own head a lot, all … then I come here and I’m still feeling blue but starting to feel like I’m growing … soon I was able to accept and feel lots of different things hence the yellow, pink and stuff …” (N10, Perpetrator)

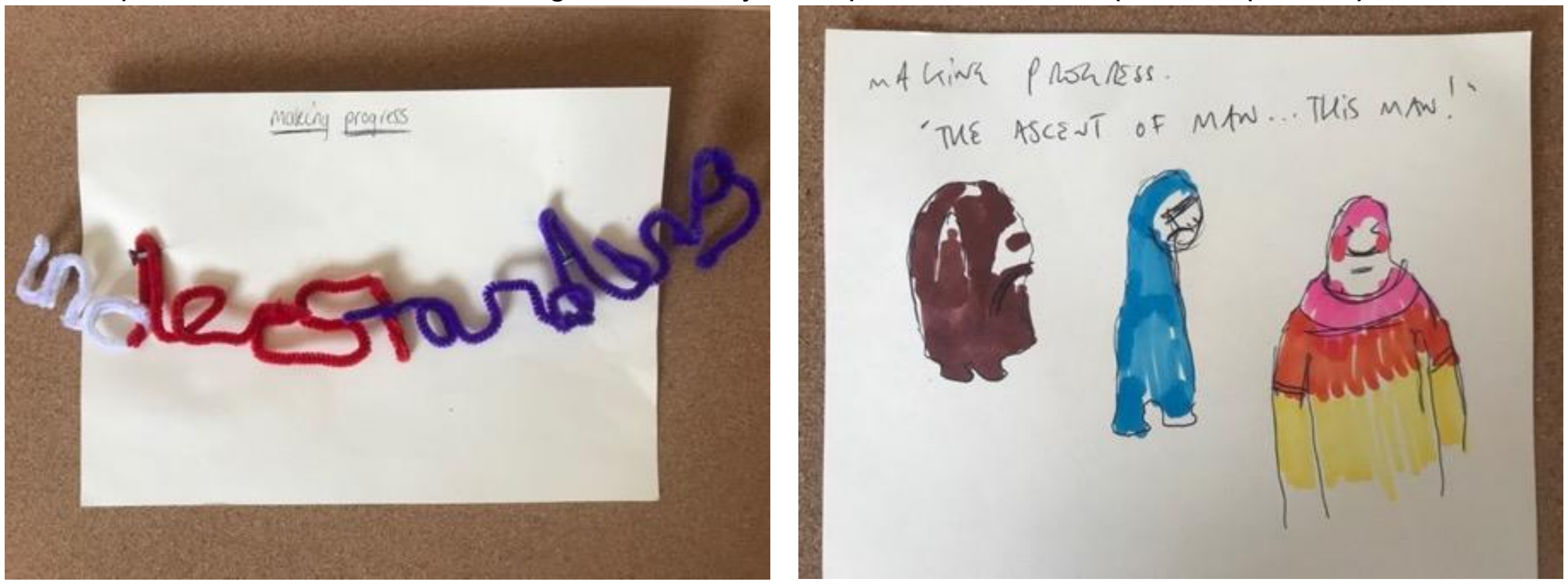

Figure 5: N1’s [Left] and N10’s [Right] fragment for “*Making Progress*”

Another common strategy was using the fragments to *challenge* false narratives of what a DVPP was, and what the programme aimed to achieve with the men. N1 (Figure 5) shared both his expected and subverted thought processes about the programme:

> “…I realised that this programme is about understanding, it’s not about punishing you, telling you off or letting you off the hook … it’s about <Stop Violence> understanding you to help you understand yourself … let them do that” (N1, Perpetrator)

### *5.1.2 Curating Identities*

The process of creating fragments of themselves for participants, with the means of sharing them with others brought larger, existential questions on their connection to abuse and violence. Several participants disclosed reflective and frequently ambivalent feelings towards the activity, both during the creative process and after they were looking over their completed set of fragments:

> “we have to share ourselves with others because of our experiences … but am I always going to be known by that experience? … when representing ourselves, when do we stop being seen as ‘perpetrators’ or, I don’t know … ‘abusers’? Are we always going to be perpetrators? Is every piece of ourselves from now representative of that?” (N2, Perpetrator)

We found this to be a compelling effect of the creative process for Group N, that many of them interpreted the process as a form of identity work around the label of perpetrator: work that is most often identified in relation to victim-survivors; whether being subject to violence makes one a victim or establishing agency in being a survivor of abuse. N2 expresses dismay that all future fragments of himself may be still ‘representative’ of his past use of violence, though not all participants were reluctant to ‘own’ this identity:

> “… we have to own up to our past, what we’ve done … if you record a mistake, you got one shot on your thoughts, we can wanna try to tape over what happened and start afresh but it don’t work like that in real life” (N9, Perpetrator)

The facilitators, while encouraging that he took responsibility for his past behaviour reminded N9 to be careful in his use of language around violence, suggesting that he should not be talking about his use of it as simply a ‘mistake’ so as to not to minimize his actions towards his victim-survivor. While the <Stop Violence> team had initially shared anxiety around whether the Group N would “*take the process seriously*” (F6, Facilitator) or “*share anything useful*” (F5, Facilitator), they both agreed that the process of having an allocated workshop where the men could share their thoughts and feelings on their identity through the fragments to be invaluable. We identified this appeared to tackle several of the challenges of unmoderated online peer support networks; a lack of initial content, sporadic engagement and disclosing unhelpful or dangerous advice.

## 6 STAGE FOUR: RECIEVING PEER SUPPORT

For the final stage of our investigation, the research team and facilitators presented the fragments from Group N to a set of new-starters to a DVPP - the Grandvens Group (Group G) - a group of eight men who were unknown to Group N. Each participant was within six weeks of starting the programme and had reportedly found the content of the previous six sessions challenging according to a facilitator (F3), identifying them as a suitable group to potentially benefit from additional support. The three-hour activity was hosted within an existing DVPP session in their local <Cloudside> hub. The final session was designed to generate progressive discourse between Group G members as they listened to and shared their experiences of interacting with the fragments provided by the Group N in order to foster the same mutually respectful practice for discussing violence. For the first half of the session, the men were split into groups of four, one set listening to a collection of fragments one-by-one (Figure 3), and the other listening to the fragments as a group before swapping halfway through. This was to compare whether the men found the peer support process most useful as an individual or a group activity. For the second half of the session, facilitators asked evaluative questions on how Group G found the process. The men were also told that if the material seen or heard through FoTP proved too distressing or emotive for the session, one of the lead facilitators (F4) could provide one-on-one behaviour change work outside of the group context to any individual who requested it. The participants were also notified that their attendance was not connected to their course evaluation, unless they were to disclose a safeguarding concern to the safety of the their victim-survivor.

### 6.1 Stage Four: Findings

We identified two prominent categories focusing on examining the role of receiving peer support from Group N from our data analysis that we discuss below: *Looking Back to Look Forward* and *Communicating Honesty*.

#### *6.1.1 Looking Back to Look Forward*

Each man in Group G could not have any direct communication with the participants of Group N due to an agreed upon ethical component of the research project (*3.1.1 Ethical Approval*). Nevertheless, even though each man did not know who had produced each fragment set, many participants shared that they "*felt connected somehow*" (G2) to some of the accounts provided by Group N through an anonymous channel of communication:

> "just knowing that other guys [are] in the same position as me, like, cared enough to give back to us … yeah that's cool man … I might not know them but in some ways it looked like I was lookin' into my future … with the letter to yourself, I'd like to get to a stage where you're asking me to do that for the next batch of guys" (G4, Perpetrator)

We found this an interesting way that G4 perceived that Group N were 'giving back' through emotional support by physically providing them their fragments. We noticed many participants shared finding the process of listening, replaying and commenting on a participant's fragments to be motivational and encouraging for their own speculative process through the behaviour change process: even imagining themselves as positions as peers themselves ('for the next batch of guys'). Some members of Group G suggested ways that the FoTP could be extended to include a solitary activity:

> "I think askin' people to make fragments as they go along, rather than right at the end of the process, you know capture that rawness of how someone feels then, instead of how they remember feeling … then I don't know, getting them to check it over at the end? Let them see if they've actually changed in themselves …" (G6, Perpetrator)

Many participants within Group G saw value in using the process of crafting fragments at different parts of the behaviour change process as providing a 'rawer' view of a journey, rather than a more positive retrospective look at the end. This was agreed by the facilitators that crafting fragments throughout the process of change could also act as a means of self-reflection and demonstrating change (or lack thereof) within a perpetrator.

### *6.1.2 Communicating Honesty*

The final stage elicited a range of responses around how authenticity or honesty was communicated through creative expression, and subsequently judged by Group G. Potentially due to the fact Group G were just starting the process, most participants spent a lot of time discussing the *First Impressions* fragments with the group over other moments. During the playback of these fragments, participants shared with the facilitators that they noticed that many of Group N had described detailed explanations of general anxiety on starting the DVPP:

> "G7: Listening to them speak and seeing the comic they represented, yeah that stuff was super depressing … a lot more real than I was expecting yeah … because you can hear them sayin' it in their own words and what they've chosen to represent about themselves … even if it's not super crafty, I think that just helps you know it's real men like us …"

> "G3: But you gotta rely on the fact we have just been listening to men who have really had their experience! It could be just <Stop Violence> having a crafting session and then putting on funny voices to make those fragments [laughs]".

This exchange between G3 and G7 directly connected the value of the fragments to two important aspects of honesty; the honesty that Group N were sharing 'authentic' accounts, and the honesty in the process of providing these accounts by the research team and facilitators. The risk of <Stop Violence> presenting 'inauthentic' or 'faked' accounts was discussed in detail, though many participants agreed that the combination of both the audio and physical creative fragments could potentially mitigate this concern:

> "I think having audio and the physical stuff both together to work as one, it helps to bring you closer to understanding that guy … but then each fragment is only connected by that little [RFID] sticker and you could lose that and lose their voice … or say that <Stop Violence> don't like the stick drawings on a comic with something more stylish … you'd never know there'd been a change." (G8, Perpetrator)

Many participants established that it was the combination of creative expression and the audio that could help ensure that the fragments were authentic, and not manipulated through the service interested in presenting a 'stylish' version of their fragments for other reasons such as an external evaluation on the service.

## 7 DISCUSSION

While research into technologies to support desistance from harmful behaviours is starting to be explored, it is less common to design for individuals who subject others to domestic abuse, despite the longitudinal support that perpetrators frequently require. Our *Fragments of the Past* project was conceived as a first step in identifying ways to sustain moderated, pro-social relations for behaviour change support beyond a programme, while seeking to overcome the ethical, practical and pragmatic challenges that can occur in traditional and online peer group creation. The ethical component of our research gave our study a particular focus that required us to identify how the process of creating novel channels of anonymous, asynchronous communication could simulate peer support. In the following, we describe how our findings can inform the design of *Channels for Passing Support*, support the process of *Identity Work Around Perpetration* and examine *Collective Responsibility for Individual Responsibility* for perpetrators and other challenging groups.

### 7.1 Channels for Passing Support

Digitally-mediated peer support networks for reducing harm are subject to a number challenges in their creation from the sparsity of engagement to the lack of accountability for users who reinforce abusive behavioural and cultural norms [44, 77]. One such challenge is the risk of collusion and escalation of risk, that is of particular concern for perpetrators in unmoderated spaces in ways that are not shared across all social groups. With the lack of resources or studies for post-programme desistance, this can make the creation of such supportive processes between perpetrators even more challenging to carefully design – even though the loss of such is clearly identified as a risk factor to the re-uptake in violent behaviours [49, 55]. While we note that *Fragments of the Past* was able to simulate aspects of a peer support process, so that members of Group G reported feeling perceived support (a valuable indicator for improved health outcomes [43]), we note that the interpersonal benefit of Group N providing peer support so that they felt valued or included was potentially absent from the process [68, 69]. The lack of back-and-forth communication between the two groups is a direct result of the ethical dimensions surrounding how this research study was conducted; as a result we realise we had to focus on designing for a meaningful one-way information transfer beyond the use of traditional digital input devices - an approach notably unpopular with our participants (*3.5.2 Mixing the Digital and Non-Digital*).

In this work we identified that the mixed-media medium of ‘fragments’ or artefacts could provide an appropriate channel for *passing* supportive peer content to other people finding taking their first steps towards desistance to be challenging. This was because artefacts took on a form of *temporal permanence* – objects that remain fixed across time – artefacts that were able to capture, express and importantly make manifest insightful reflections on emotional support to be physically passed between groups. While we acknowledge that other studies that we have included in this work draw attention to this ‘fixedness’ quality of artefacts for remembrance [33, 79] and gift-giving [74], we found that this permanence provided the means to capture both a representation of change for the provider (e.g. N6 in *5.1.1 Audibly Augmenting Reality*) and provide motivation for change for the receiver (e.g. *6.1.1 Looking Back to Look Forward*). We are the first to explicitly highlight this powerful dual role that mixed-media artefacts can play within the creation of peer support activities, and encourage other designers, practitioners and researchers to actively pay explicit attention to how relational and communicative dynamics can play out through digital artefacts where synchronous communication is inaccessible. A potential way that future research could seek to innovate further is through exploring how artefacts can continuously be passed between groups to observe how fragment creators see how their artefacts are received.

### 7.2 Identity Work Around Perpetration

We identified that through the process of making many of our participants engaged in work around their own identity with respect to the label ‘perpetrator’; whether seeking to reject it or accept it (*3.5.1 Balancing Parts with Wholes; 5.1.2 Curating Identities*). Bringing out such complex and challenging questions in relation to harm, past memory, and identity adds to the ever-growing corpus of work within HCI that understands the value in further exploring the process of making to communicate sensitive and challenging topics [18, 48, 88]. We note that due to the flexibility in the process of using JigsAudio as a medium for the co-designed process of fragments of the past, this also permitted our participants to share only what they felt comfortable sharing about their identities, behaviours and how these may have subsequently changed across the behaviour change process. As such we suggest that when creating both asynchronous and synchronous communicative groups, that such practitioners and researchers might investigate including mechanisms for creative expression to assist people in communicating closely held feelings and experiences. We do so to encourage cultivating approaches in HCI that seek to uncover greater emotional understandings of the growing participant base that the field works with to lead to improved health and wellbeing outcomes, and for the protection of vulnerable groups.

We acknowledge that caution must be applied to repeating such a process where participants who have used harm are not free to present themselves in such a light where their abusive behaviours are legitimised by the minimisation, denial or blame of others [5, 35]. Providing those with the ability to craft an identity, particularly people who may have carefully curated an acceptable public-facing identity so as to hide their abuse of their partners, is a particular underlined risk with this social group. As each stage was carefully co-planned with the <Stop Violence> facilitators, support processes that could potentially undermine the messages of a DVPP through negative feedback loops were carefully caught and challenged before material could be added to the support process. In such a way, we acknowledge that our process could lead to tension between an authentic yet unhelpful representation of emotive support, and a moderated (thereby slightly inauthentic) yet useful representation. For future work we anticipate exploring how both ‘helpful’ and ‘unhelpful’ fragments can be used to engage perpetrators in independent, individual reflection on prior thoughts and behaviours (as suggested by G6), and how these might compare to existing in-person reflective evaluations such as mid-programme reports, or post-programme risk assessments. Such an approach that we have described within this work depicts a relatively low-cost (from an organizational perspective) means of revisiting these manifestations and providing these impressions to others. While we identified our process illuminated interesting findings for perpetrators, we would be keen to explore whether the same process of catching negative feedback loops – loops that do indeed impact on other social groups might be adapted to that context.

### 7.3 Collective Responsibility for Individual Responsibility

It can be deeply uncomfortable to acknowledge that the process of sustainable behaviour change is rarely a linear movement away from violence, by rather a "*dynamic pathway*" of moments of relapse, confusion and resistance to change [49]. We acknowledge that our ethical approval that shaped the way this research was conducted most likely stemmed from a concern for protecting vulnerable population groups from undue harm, a concern that we seek to continuously respond to, and be informed by, across our work with perpetrators [28]. Nevertheless, a decision of non-communication, from the perspective of our support organisation, does place undue focus on isolated, static and (arguably) speculative moments of our participants' journey to non-violence where they could behave irresponsibly. This is all the more reason, in line with Bellini et al.'s calls in [6], to better understand the use of violence as a behaviour subject to internal and external factors, rather than rooted in a normative impression of who or what a 'perpetrator' is and what actions they might perform.

While we are not advocating to disregard the possibility that some participants might behave in ways that may undermine the messages of the behaviour change course, this is part of the learning process inherent to DVPPs, and that group accountability is the very source of pro-social enforcement [26]. Our participants were not afforded the "*spaces for negotiation*" to take responsibility for others, even within a safety-focused and heavily moderated specialist environment such as <Stop Violence> where collusion and risk are continuously and rigorously evaluated for [6]. Our investigation into developing responsible interaction frameworks sought to not only strongly encourage perpetrators in *taking* responsibility for their violence, but also examine how <Stop Violence> could permit *giving* the perpetrators responsibility to do so. Such an exchange can rely on significant bonds of trust and honesty between both groups of which our participants were more than well aware in openly discussing; such as the dishonesty by the men in a misleading representation of themselves through their fragments, or through the facilitators producing inauthentic fragments.

The chance for risk of entire groups going into relapse or partaking in the reinforcement of negative patterns appears out of step with the greater possibility for the collectives to regulate when outliers occur [47, 67], particularly since responsible behaviour is determined dynamically overtime rather than within a single incident. In this way, such blanket verdicts may make ethical clearance and engagement with such groups more challenging, and can lead to a lack of evidence for informing the safe practices that are necessary to reduce the harm to vulnerable population groups [28, 57]. In such a way, despite the ethical dimensions that working with complex groups can incur, researchers should not be intimidated to, in the words of Brown et al. "*ground [their] sensitivities of those being based on everyday practice and judgement*" [12] as researchers within HCI strive to behave ethically and do ethics within an ever expanding field.

## 8 CONCLUSION

In this paper we outlined the benefits that peer support mechanisms can provide to individuals following behaviour change interventions for harmful behaviour and the challenges encountered with the creation and sustainability of networks. We worked directly with the challenges of direct inner-personal communication through detailing the design, deployment and evaluation of Fragments of the Past, an asynchronous peer support process between two groups of current and former perpetrators of domestic violence. During our subsequent deployment, we demonstrate that our system was able to simulate specific aspects of translating peer support for the receivers, still providing the givers with the ability to perform identity work in relation to the topic of their use of violence. We conclude with an overview of design suggestions for how HCI researchers and practitioners might mobilise such an approach in other sensitive settings.

## 9 ACKNOWLEDGEMENTS

We would like to thank all of our participants, partners and reviewers for their contribution to, and their on-going support in this research study. We especially extend this appreciation to our shepherd who graciously gave up their time on the run up to the holiday period to assist us in better communicating the ethical dimensions of this work. This research was funded through the EPSRC Centre for Doctoral Training in Digital Civics (EP/L016176/1). Data supporting this publication is not openly available due to confidentiality considerations. Access may be possible under appropriate agreement. Additional metadata record at https://doi.org/10.25405/data.ncl.13415015

## 10 REFERENCES

[1] Adhia, A. et al. 2020. Life experiences associated with change in perpetration of domestic violence. *Injury Epidemiology*. 7, 1 (Aug. 2020), 37. DOI:https://doi.org/10.1186/s40621-020-00264-z.

[2] Banyard, V.L. 2011. Who will help prevent sexual violence: Creating an ecological model of bystander intervention. *Psychology of Violence*. 1, 3 (2011), 216–229. DOI:https://doi.org/10.1037/a0023739.

[3] Barrett, A.K. and Murphy, M.M. 2020. Feeling Supported in Addiction Recovery: Comparing Face-to-Face and Videoconferencing 12-Step Meetings. *Western Journal of Communication*. 0, 0 (Jul. 2020), 1–24. DOI:https://doi.org/10.1080/10570314.2020.1786598.

[4] Baumer, E.P.S. and Silberman, M.S. 2011. When the Implication is Not to Design (Technology). *Proceedings of the SIGCHI Conference on Human Factors in Computing Systems* (New York, NY, USA, 2011), 2271–2274.

[5] Bellini, R. et al. 2020. Choice-Point: Fostering Awareness and Choice with Perpetrators in Domestic Violence Interventions. *Proceedings of the 2020 CHI Conference on Human Factors in Computing Systems* (New York, NY, USA, Apr. 2020), 1–14.

[6] Bellini, R. et al. 2020. Mechanisms of Moral Responsibility: Rethinking Technologies for Domestic Violence Prevention Work. *Proceedings of the 2020 CHI Conference on Human Factors in Computing Systems* (New York, NY, USA, Apr. 2020), 1–13.

[7] Bellini, R. et al. 2018. “That Really Pushes My Buttons”: Designing Bullying and Harassment Training for the Workplace. *Proceedings of the 2018 CHI Conference on Human Factors in Computing Systems* (New York, NY, USA, 2018), 235:1-235:14.

[8] Bellini, R. et al. 2019. Vocalising Violence: Using Violent Mens’ Voices for Service Delivery and Feedback. *Proceedings of the 9th International Conference on Communities & Technologies - Transforming Communities* (New York, NY, USA, 2019), 210–217.

[9] Bennett, G.G. and Glasgow, R.E. 2009. The Delivery of Public Health Interventions via the Internet: Actualizing Their Potential. *Annual Review of Public Health*. 30, 1 (2009), 273–292. DOI:https://doi.org/10.1146/annurev.publhealth.031308.100235.

[10] Birbeck, N. et al. 2017. Self Harmony: Rethinking Hackathons to Design and Critique Digital Technologies for Those Affected by Self-Harm. *Proceedings of the 2017 CHI Conference on Human Factors in Computing Systems* (Denver, Colorado, USA, May 2017), 146–157.
[11] Braithwaite, D.O. et al. 1999. Communication of social support in computer-mediated groups for people with disabilities. *Health Communication*. 11, 2 (1999), 123–151. DOI:https://doi.org/10.1207/s15327027hc1102_2.
[12] Brown, B. et al. 2016. Five Provocations for Ethical HCI Research. *Proceedings of the 2016 CHI Conference on Human Factors in Computing Systems* (New York, NY, USA, 2016), 852–863.
[13] Bushway, S.D. et al. 2001. An Empirical Framework for Studying Desistance as a Process*. *Criminology*. 39, 2 (May 2001), 491–516. DOI:https://doi.org/10.1111/j.1745-9125.2001.tb00931.x.
[14] Campbell, R. 2013. *Emotionally Involved: The Impact of Researching Rape*. Routledge.
[15] Cannon, C. et al. 2020. The Pursuit of Research-supported Treatment in Batterer Intervention: The Role of Professional Licensure and Theoretical Orientation for Duluth and CBT Programs. *Journal of Evidence-Based Social Work*. 17, 4 (Jul. 2020), 469–485. DOI:https://doi.org/10.1080/26408066.2020.1775744.
[16] Chancellor, S. et al. 2018. Norms Matter: Contrasting Social Support Around Behavior Change in Online Weight Loss Communities. *Proceedings of the 2018 CHI Conference on Human Factors in Computing Systems* (Montreal QC, Canada, Apr. 2018), 1–14.
[17] Chayn 2020. Projects. *Chayn*.
[18] Clarke, R. et al. 2013. Digital Portraits: Photo-sharing After Domestic Violence. *Proceedings of the SIGCHI Conference on Human Factors in Computing Systems* (New York, NY, USA, 2013), 2517–2526.
[19] Cohen, S. 2004. Social Relationships and Health. *American Psychologist*. 59, 8 (Nov. 2004), 676–684.
[20] Congress, E.F. for M.I.I. 2009. *Medical Informatics in a United and Healthy Europe: Proceedings of MIE 2009*. IOS Press.
[21] Cupchik, G. 2001. Constructivist Realism: An Ontology That Encompasses Positivist and Constructivist Approaches to the Social Sciences. *Forum Qualitative Sozialforschung / Forum: Qualitative Social Research*. 2, 1 (Feb. 2001). DOI:https://doi.org/10.17169/fqs-2.1.968.
[22] Dalley, T. 1984. *Art as Therapy: An Introduction to the Use of Art as a Therapeutic Technique*. Routledge.
[23] Daniels, A.S. et al. 2012. Pillars of peer support: advancing the role of peer support specialists in promoting recovery. *The Journal of Mental Health Training, Education and Practice*. 7, 2 (Jan. 2012), 60–69. DOI:https://doi.org/10.1108/17556221211236457.
[24] Dash Risk Checklist – Saving lives through early risk identification, intervention and prevention: *https://www.dashriskchecklist.co.uk/*. Accessed: 2017-12-11.
[25] DeKeseredy, W.S. and Schwartz, M.D. 2013. *Male Peer Support and Violence Against Women: The History and Verification of a Theory*. Northeastern University Press.
[26] Domestic violence perpetrator programmes:an historical overview.: 2013. *https://www.dur.ac.uk/resources/criva/anhistoricaloverviewbriefingnote.pdf*. Accessed: 2018-05-25.
[27] Donovan, C. and Griffiths, S. 2015. Domestic Violence and Voluntary Perpetrator Programmes: Engaging Men in the Pre-Commencement Phase. *The British Journal of Social Work*. 45, 4 (Jun. 2015), 1155–1171. DOI:https://doi.org/10.1093/bjsw/bct182.
[28] Downes, J. et al. 2014. Ethics in Violence and Abuse Research - a Positive Empowerment Approach: *Sociological Research Online*. (Mar. 2014). DOI:https://doi.org/10.5153/sro.3140.
[29] Durrant, A.C. et al. 2014. Human values in curating a human rights media archive. *Proceedings of the SIGCHI Conference on Human Factors in Computing Systems* (New York, NY, USA, Apr. 2014), 2685–2694.
[30] Farrall, S. 2004. Social capital and offender reintergration: making probation desistance focused. *After Crime and Punishment: Pathways to Offender Reintegration*. Routledge.

[31] Frohlich, D. and Murphy, R. 2000. The Memory Box. *Personal and Ubiquitous Computing*. 4, 4 (Jan. 2000), 238–240. DOI:https://doi.org/10.1007/PL00000011.
[32] Glaser, B. and Strauss, A. 1999. *The Discovery of Grounded Theory: Strategies for Qualitative Research*. Routledge.
[33] Goodman, E. and Rosner, D. 2011. From garments to gardens: negotiating material relationships online and "by hand." *Proceedings of the SIGCHI Conference on Human Factors in Computing Systems* (New York, NY, USA, May 2011), 2257–2266.
[34] Hamilton, L. et al. 2013. Domestic Violence Perpetrator Programs in Europe, Part I: A survey of Current Practice. *International Journal of Offender Therapy and Comparative Criminology*. 57, 10 (Oct. 2013), 1189–1205. DOI:https://doi.org/10.1177/0306624X12469506.
[35] Hearn, J.R. 1998. *The Violences of Men: How Men Talk about and How Agencies Respond to Men's Violence to Women*. Sage Publications.
[36] Hegarty, K. et al. 2016. Interventions to support recovery after domestic and sexual violence in primary care. *International Review of Psychiatry*. 28, 5 (Sep. 2016), 519–532. DOI:https://doi.org/10.1080/09540261.2016.1210103.
[37] Hester, M. et al. 2006. Domestic Violence Perpetrators: Identifying Needs to Inform Early Intervention. (2006).
[38] Heyer, J. et al. 2020. Opportunities for Enhancing Access and Efficacy of Peer Sponsorship in Substance Use Disorder Recovery. *Proceedings of the 2020 CHI Conference on Human Factors in Computing Systems* (New York, NY, USA, Apr. 2020), 1–14.
[39] Hicks, C.M. et al. 2016. Framing Feedback : Choosing Review Environment Features that Support High Quality Peer Assessment. (2016).
[40] Jamieson, S. and Mikko Vesala, K. 2008. Exploring Men's Perpetrator Programs In Small Rural Communities. *Rural Society*. 18, 1 (Apr. 2008), 39–50. DOI:https://doi.org/10.5172/rsj.351.18.1.39.
[41] Kalanithi, J.J. and Bove, V.M. 2008. Connectibles: tangible social networks. *Proceedings of the 2nd international conference on Tangible and embedded interaction* (New York, NY, USA, Feb. 2008), 199–206.
[42] Kelly, L. and Westmarland, N. 2015. *Domestic Violence Perpetrator Programmes: Steps Towards Change. Project Mirabal Final Report.* Durham University.
[43] Kim, H. et al. 2017. Complementary Support from Facilitators and Peers for Promoting mHealth Engagement and Weight Loss. *Journal of Health Communication*. 22, 11 (Nov. 2017), 905–912. DOI:https://doi.org/10.1080/10810730.2017.1373876.
[44] Kushner, T. and Sharma, A. 2020. Bursts of Activity: Temporal Patterns of Help-Seeking and Support in Online Mental Health Forums. *Proceedings of The Web Conference 2020* (New York, NY, USA, Apr. 2020), 2906–2912.
[45] Lichtman, M. *Qualitative Research in Education: A User's Guide*.
[46] Lindsay, S. et al. 2012. Empathy, participatory design and people with dementia. *Proceedings of the SIGCHI Conference on Human Factors in Computing Systems* (Austin, Texas, USA, May 2012), 521–530.
[47] Maitland, J. and Chalmers, M. 2011. Designing for peer involvement in weight management. *Proceedings of the SIGCHI Conference on Human Factors in Computing Systems* (New York, NY, USA, May 2011), 315–324.
[48] Marshall, K. et al. 2014. Making wellbeing: a process of user-centered design. *Proceedings of the 2014 conference on Designing interactive systems* (New York, NY, USA, Jun. 2014), 755–764.
[49] McNeill, F. and Maruna, S. 2007. Giving up and giving back: desistance, generativity and social work with offenders. *Developments in Social Work with Offenders*. Jessica Kingsley Publishers. 226.

[50] Mead, S. et al. 2001. Peer support: a theoretical perspective. *Psychiatric Rehabilitation Journal*. 25, 2 (2001), 134–141. DOI:https://doi.org/10.1037/h0095032.
[51] Mey, G. and Dietrich, M. 2016. From Text to Image—Shaping a Visual Grounded Theory Methodology. *Forum Qualitative Sozialforschung / Forum: Qualitative Social Research*. 17, 2 (Apr. 2016). DOI:https://doi.org/10.17169/fqs-17.2.2535.
[52] Mo, P.K.H. and Coulson, N.S. 2014. Are online support groups always beneficial? A qualitative exploration of the empowering and disempowering processes of participation within HIV/AIDS-related online support groups. *International Journal of Nursing Studies*. 51, 7 (Jul. 2014), 983–993. DOI:https://doi.org/10.1016/j.ijnurstu.2013.11.006.
[53] Moncur, W. et al. 2015. Story Shell: the participatory design of a bespoke digital memorial. *Proceedings of the 4th Participatory Innovation Conference PIN-C 2015* (May 2015), 470–477.
[54] Morran, D. 2011. Re-education or recovery? Re-thinking some aspects of domestic violence perpetrator programmes. *Probation Journal*. 58, 1 (Mar. 2011), 23–36. DOI:https://doi.org/10.1177/0264550510388968.
[55] Morran, D. 2006. Thinking Outside the Box: Looking Beyond Programme Integrity: The Experience of a Domestic Violence Offenders Programme. *British Journal of Community Justice*. 4, 1 (2006), 7–18.
[56] Mugellini, E. et al. 2007. Using personal objects as tangible interfaces for memory recollection and sharing. *Proceedings of the 1st international conference on Tangible and embedded interaction* (New York, NY, USA, Feb. 2007), 231–238.
[57] Newman, E. et al. 1999. Assessing the ethical costs and benefits of trauma-focused research. *General Hospital Psychiatry*. 21, 3 (Jun. 1999), 187–196. DOI:https://doi.org/10.1016/s0163-8343(99)00011-0.
[58] Nicholson, J. and McGlasson, J. 2020. CyberGuardians: Improving Community Cyber Resilience Through Embedded Peer-to-Peer Support. *Companion Publication of the 2020 ACM Designing Interactive Systems Conference* (New York, NY, USA, Jul. 2020), 117–121.
[59] O'Leary, K. et al. 2017. Design Opportunities for Mental Health Peer Support Technologies. *Proceedings of the 2017 ACM Conference on Computer Supported Cooperative Work and Social Computing* (New York, NY, USA, Feb. 2017), 1470–1484.
[60] O'Leary, K. et al. 2018. “Suddenly, we got to become therapists for each other”: Designing Peer Support Chats for Mental Health. *Proceedings of the 2018 CHI Conference on Human Factors in Computing Systems* (New York, NY, USA, Apr. 2018), 1–14.
[61] Pang, C.E. et al. 2013. Technology preferences and routines for sharing health information during the treatment of a chronic illness. *Proceedings of the SIGCHI Conference on Human Factors in Computing Systems* (New York, NY, USA, Apr. 2013), 1759–1768.
[62] Pater, J. and Mynatt, E. 2017. Defining Digital Self-Harm. *Proceedings of the 2017 ACM Conference on Computer Supported Cooperative Work and Social Computing* (Portland, Oregon, USA, Feb. 2017), 1501–1513.
[63] Pearson, D.A.S. and Ford, A. 2018. Design of the "Up2U" domestic abuse perpetrator programme. *Journal of Aggression, Conflict and Peace Research*. 10, 3 (Jan. 2018), 189–201. DOI:https://doi.org/10.1108/JACPR-04-2017-0280.
[64] Penney, D. 2018. *Defining "Peer Support": Implications for Policy, Practice, and Research*. Technical Report #1. Advocates for Human Potential.
[65] Rainey, J. et al. 2019. Gabber: Supporting Voice in Participatory Qualitative Practices. *CHI Conference on Human Factors in Computing Systems Proceedings (CHI 2019)*. (May 2019). DOI:https://doi.org/https://doi.org/10.1145/3290605.3300607.
[66] Respect UK 2020. *A Domestic Abuse Perpetrator Strategy for England And Wales*. Respect.

[67] Rubya, S. and Yarosh, S. 2017. Video-Mediated Peer Support in an Online Community for Recovery from Substance Use Disorders. *Proceedings of the 2017 ACM Conference on Computer Supported Cooperative Work and Social Computing* (New York, NY, USA, Feb. 2017), 1454–1469.

[68] Satinsky, E.N. et al. 2020. Adapting a peer recovery coach-delivered behavioral activation intervention for problematic substance use in a medically underserved community in Baltimore City. *PLOS ONE*. 15, 1 (Jan. 2020), e0228084. DOI:https://doi.org/10.1371/journal.pone.0228084.

[69] Schildkraut, J. et al. 2020. The Survivor Network: The Role of Shared Experiences in Mass Shootings Recovery. *Victims & Offenders*. 0, 0 (May 2020), 1–30. DOI:https://doi.org/10.1080/15564886.2020.1764426.

[70] Schmid, T. 2005. *Promoting Health Through Creativity: For professionals in health, arts and education*. John Wiley & Sons.

[71] Schueller, S.M. et al. 2019. Use of Digital Mental Health for Marginalized and Underserved Populations. *Current Treatment Options in Psychiatry*. (2019). DOI:https://doi.org/10.1007/s40501-019-00181-z.

[72] Scourfield, J.B. and Dobash, R.P. 1999. Programmes for Violent Men: Recent Developments in the UK. *The Howard Journal of Criminal Justice*. 38, 2 (1999), 128–143. DOI:https://doi.org/10.1111/1468-2311.00122.

[73] Simonsen, J. ed. 2013. *Routledge International Handbook of Participatory Design*. Routledge.

[74] Spence, J. 2019. Inalienability: Understanding Digital Gifts. *Proceedings of the 2019 CHI Conference on Human Factors in Computing Systems - CHI '19* (Glasgow, Scotland Uk, 2019), 1–12.

[75] Strohmayer, A. et al. 2015. Exploring Learning Ecologies among People Experiencing Homelessness. *Proceedings of the 33rd Annual ACM Conference on Human Factors in Computing Systems* (New York, NY, USA, Apr. 2015), 2275–2284.

[76] Thieme, A. et al. 2013. Design to Promote Mindfulness Practice and Sense of Self for Vulnerable Women in Secure Hospital Services. *Proceedings of the SIGCHI Conference on Human Factors in Computing Systems* (New York, NY, USA, 2013), 2647–2656.

[77] Tseng, E. et al. 2020. The Tools and Tactics Used in Intimate Partner Surveillance: An Analysis of Online Infidelity Forums. *arXiv:2005.14341 [cs]*. (May 2020).

[78] Walker, K. et al. 2013. Desistance from intimate partner violence: A critical review. *Aggression and Violent Behavior*. 18, 2 (Mar. 2013), 271–280. DOI:https://doi.org/10.1016/j.avb.2012.11.019.

[79] Wallace, J. et al. 2013. Making design probes work. *Proceedings of the SIGCHI Conference on Human Factors in Computing Systems* (New York, NY, USA, Apr. 2013), 3441–3450.

[80] Wei, H. et al. 2019. MemoryReel: A Purpose-designed Device for Recording Digitally Connected Special Moments for Later Recall and Reminiscence. *Proceedings of the Thirteenth International Conference on Tangible, Embedded, and Embodied Interaction* (New York, NY, USA, Mar. 2019), 135–144.

[81] Westmarland, N. and Bows, H. 2018. *Researching Gender, Violence and Abuse: Theory, Methods, Action*. Routledge.

[82] Westmarland, N. and Kelly, L. 2013. Why Extending Measurements of 'Success' in Domestic Violence Perpetrator Programmes Matters for Social Work. *The British Journal of Social Work*. 43, 6 (Sep. 2013), 1092–1110. DOI:https://doi.org/10.1093/bjsw/bcs049.

[83] Wilson, A. and Tewdwr-Jones, M. 2019. Let's draw and talk about urban change: Deploying digital technology to encourage citizen participation in urban planning. *Environment and Planning B: Urban Analytics and City Science*. (Feb. 2019), 2399808319831290. DOI:https://doi.org/10.1177/2399808319831290.

[84] Wilson, R.J. et al. 2009. Circles of Support & Accountability: a Canadian national replication of outcome findings. *Sexual Abuse: A Journal of Research and Treatment*. 21, 4 (Dec. 2009), 412–430. DOI:https://doi.org/10.1177/1079063209347724.

[85] Wright, E.M. 2015. The Relationship Between Social Support and Intimate Partner Violence in Neighborhood Context. *Crime & Delinquency*. 61, 10 (Dec. 2015), 1333–1359. DOI:https://doi.org/10.1177/0011128712466890.

[86] Yarosh, S. 2013. Shifting Dynamics or Breaking Sacred Traditions?: The Role of Technology in Twelve-step Fellowships. *Proceedings of the SIGCHI Conference on Human Factors in Computing Systems* (New York, NY, USA, 2013), 3413–3422.

[87] Zimmerman, J. et al. 2008. New Methods for the Design of Products That Support Social Role Transitions. *Artifact*. 2, 3–4 (Sep. 2008), 190–206. DOI:https://doi.org/10.1080/17493460802527113.

[88] Stepping Out of The Box Art Exhibition. *Changing Relations*.